\documentclass[english,aps,preprint]{revtex4}
\usepackage[T1]{fontenc}
\usepackage[latin9]{inputenc}
\usepackage{graphicx}
\usepackage{amssymb}

\makeatletter
\providecommand{\tabularnewline}{\\}

\@ifundefined{textcolor}{} {
 \definecolor{BLACK}{gray}{0}
 \definecolor{WHITE}{gray}{1}
 \definecolor{RED}{rgb}{1,0,0}
 \definecolor{GREEN}{rgb}{0,1,0}
 \definecolor{BLUE}{rgb}{0,0,1}
 \definecolor{CYAN}{cmyk}{1,0,0,0}
 \definecolor{MAGENTA}{cmyk}{0,1,0,0}
 \definecolor{YELLOW}{cmyk}{0,0,1,0}
 }

\makeatother

\usepackage{babel}

\begin{document}

\title{Forward-Backward Asymmetries of Fourth Family Fermions Through the
$Z'$ Models at Linear Colliders}

\author{V. Ar{\i}}

\email{volkan.ari@science.ankara.edu.tr}

\affiliation{Ankara University, Faculty of Sciences, Department of Physics, 06100,
Tandogan, Ankara.}

\author{O. \c{C}ak{\i}r}

\email{ocakir@science.ankara.edu.tr}

\affiliation{Ankara University, Faculty of Sciences, Department of Physics, 06100,
Tandogan, Ankara.}

\author{V. \c{C}etinkaya}

\email{volkan.cetinkaya@science.ankara.edu.tr}

\affiliation{Ankara University, Faculty of Sciences, Department of Physics, 06100,
Tandogan, Ankara.}

\begin{abstract}
We investigate the forward-backward asymmetries in the pair production
of fourth family fermions through the new $Z'$ interactions in the
$e^{+}e^{-}$ collisions. The $Z'$ boson having family universal
couplings can contribute to the pair production of fourth family fermions
via the $s$-channel exchange. The linear colliders will provide a
clean environment for the physics of $Z'$ boson to measure its couplings
precisely. The effects of the $Z'$ boson to the asymmetries are shown
to be important in some parameter regions for different $Z'$ models.
Among these parameters, the invariant mass distribution $m_{F\bar{F}}$
will be an important measurement to constrain the $Z'$ models. Providing
the fourth family fermion exist in an accesssible mass range, a $\chi^{2}$
analysis can be used to probe the $Z'$ models at linear colliders.
\end{abstract}

\maketitle

\section{Introduction}

Even though we observe three families of quarks and leptons of the
Standard Model (SM), however there could be a fourth family if their
masses and mixings are beyond our present experimental reach. A family extension
to the SM fermion families contains the quarks $t'$ and $b'$, and
charged lepton $l'$ with its associated neutrino $\nu'$. The allowed
parameter space for a fourth family is restricted by the experimental
searches, precision electroweak measurements, theoretical constraints
from the requirements of unitarity and perturbativity. 

Recent searches at Large Hadron Collider (LHC) experiments have considered pair 
production of the fourth family quarks. The data of 1 fb$^{-1}$ from 
ATLAS experiment at the LHC (7 TeV) restricts the 
masses of $t'$ and $b'$ quarks: $m_{t'}>404$ GeV at $95\%$ CL. \cite{ATLAS12_1} 
assuming $t'\to W^+b$ and $m_{b'}>450$ GeV at $95\%$ CL. \cite{ATLAS12_2} 
assuming $b'\to W^-t$ and taking into account 
the subsequent decays into same-sign dilepton final state. A search 
for pair produced bottom-like quarks ($b'$) in the lepton+jets channel 
by the ATLAS Collaboration excludes a $b'$ quark mass 
of less than 480 GeV \cite{ATLAS12_2}. 

Using a data sample corresponding to an integrated luminosity of 4.9 fb$^{-1}$ with 
the CMS detector at the LHC, the most stringent limits exclude the existence 
of a down-type (up-type) fourth family quark 
with masses below 611 GeV (565 GeV) \cite{CMS12_1,CMS12_2} 
assuming a branching fraction of 100$\%$ for the decays $b'\to W^-t$ and $t'\to W^+b$.

From direct production searches at LEPII, 
there is a lower limit of the order of 100 GeV
for the fourth family charged lepton and unstable neutrino. The precision
measurements restrict the mass splitting between the fourth family
leptons $|M_{l'}-M_{\nu'}|\approx30-60$ GeV and the fourth family
quarks $|M_{t'}-M_{b'}|\approx50-70$ GeV \cite{Kribs07,Erler10}.

The fourth family quarks and leptons could also couple to an extra
neutral gauge boson different from the three SM families. A new neutral
gauge boson $Z'$ can have family universal or non-universal couplings
to fermions. The indirect searches of the $Z'$ boson can also be
performed at linear colliders where the discovery limits are related
to the deviations from the SM predictions for the cross sections and
asymmetries due to the interference effects between the propagators.
The $Z'$ boson having family universal couplings can contribute to
the pair production of fourth family fermions via the $s$-channel
exchange. The linear collider provides a clean environment for $Z'$
physics, and can measure the couplings precisely.

In the extensions of the SM with $U(1)_{\psi}\times U(1)_{\chi}$
gauge symmetry, the fields $Z'_{\psi}$ and $Z'_{\chi}$ can be massive
and their states can mix, therefore, a relatively lighter mass eigenstate
can be written as $Z'(\theta)=Z'_{\psi}\cos\theta+Z'_{\chi}\sin\theta$.
A set of $Z'$ models have some special names: the sequential $Z'_{S}$
model has the same coupling to the fermions as that of the $Z$ boson
of the SM; the $Z'_{\psi}$, $Z'_{\chi}$ and $Z'_{\eta}$ models
corresponding to the specific values of the mixing angle $\theta$
($0$, $\pi/2$ and $\arctan\sqrt{3/5}$, respectively) in the $E_{6}$
model have different couplings to the fermions; the $Z'_{B-L}$ model
has the couplings related to the minimal $B-L$ (where $B$ and $L$
are baryon and lepton numbers, respectively) extension of the SM.
The detailed descriptions of the $Z'$ models, as well as the specific
references can be found in Refs. \cite{Hewett89,Leike99,Rizzo06,Langacker08}.

The Tevatron experiments excluded the sequential $Z'_{S}$ boson with 
a mass lower than $1$ TeV at $95\%$ CL. \cite{CDF_PRL11}. Recent measurements 
by the ATLAS and CMS Collaborations based on the data of 1 fb$^{-1}$ excludes 
a $Z'_{S}$ with mass lower than 1.83 TeV \cite{ATLAS11} and 1.94 TeV \cite{CMS11}, 
respectively. With the data corresponding to an integrated luminosity of 5 fb$^{-1}$ 
recorded by the ATLAS experiment, a lower limit of 2.21 TeV on the mass of sequential 
$Z'$ boson has been set at 95$\%$ CL \cite{ATLAS_CONF12}.

These limits on the $Z'$ boson mass favors high energy ($\geq1$ TeV) collisions for
the observation of signal from most of the $Z'$ models.
It is also possible that the $Z'$ bosons can be much heavy or weak enough to escape 
beyond the discovery reach expected at the LHC. 
In this case, only the indirect signatures
of $Z'$ exchanges may occur at the high energy colliders.

Recently, D0 and CDF Collaborations have measured the forward-backward
(FB) asymmetries of top quark $A_{FB}^{t}$ at Tevatron \cite{D008,CDF11}
in the large $t\bar{t}$ invariant mass region, while the $A_{FB}^{b}$
was measured in the $Z$ boson decays at LEP \cite{LEP06}, which
differ by about $3\sigma$ deviations from the SM expectations, without
affecting significantly the well behaved total cross sections. Several
models of new physics have been considered to explain these asymmetries
(see Refs. \cite{Djouadi07,Djouadi11,Berger11} and references therein)
at hadron colliders. 

In this study, we investigate the forward-backward asymmetries $A_{FB}^{F}$
for pair production of the fourth family fermions $F_{i}$ ($t',b',l',\nu'$)
within the $Z'$ models at linear collider energies of 1 TeV and 3
TeV. The linear colliders, namely the International Linear Collider
(ILC) described in \cite{Brau07,Brau071} and the Compact Linear Collider
(CLIC) described in \cite{Assmann00,Accomando04}, have been designed to meet
the baseline requirements for the planned physics programs. 
The effects of the $Z'$ boson
to the asymmetries of fourth family fermions at linear colliders are
shown to be important in some parameter regions for the sequential
model, some special $E_6$ models and the $B-L$ model. We will typically
consider the mass of the fourth family charged lepton greater than
200 GeV and the masses of the fourth family quarks greater than 600
GeV, which are safely above the direct production bounds.

\section{Interactions With Fourth Family Fermions}

The interactions of the fourth family quarks ($Q_{i}$) via neutral
gauge bosons ($g,\gamma,Z,Z'$) and fourth family leptons ($L_{i}$)
via electroweak gauge bosons ($\gamma,Z,Z'$) can be described by
the following Lagrangian. We also include the interactions of fourth
family fermions ($F_{i}$) with three known families of fermions ($f_{i}$)
through the charged currents (via $W^{\pm}$ bosons) to be read as

\begin{eqnarray}
L' & = & -g_{s}\overline{Q}_{i}T^{a}\gamma^{\mu}Q_{i}G_{\mu}^{a}-g_{e}Q_{F}\overline{F}_{i}\gamma^{\mu}F_{i}A_{\mu}-\frac{g}{2\sqrt{2}}V_{ij}\overline{F}_{i}\gamma^{\mu}(1-\gamma^{5})f_{j}W_{\mu}\nonumber \\
 &  & -\frac{g_{Z}}{2}\overline{F}_{i}\gamma^{\mu}(C_{V}^{F}-C_{A}^{F}\gamma^{5})F_{i}Z_{\mu}-\frac{g_{Z'}}{2}\overline{F}_{i}\gamma^{\mu}(C_{V}^{'F}-C_{A}^{'F}\gamma^{5})F_{i}Z'_{\mu}+\mbox{H.c.}\label{eq:1}
\end{eqnarray}
where $g_{s}$, $g_{e}$, $g_{Z}$ are the strong, electromagnetic
and weak-neutral coupling constants, respectively. The $G_{\mu}^{a}$,
$A_{\mu}$, $W_{\mu}$ and $Z_{\mu}$ are the fields for gluons, photon,
$W$ and $Z$ bosons, respectively. The $C_{V}^{'}$ ($C_{V}$) and
$C_{A}^{'}$ ($C_{A}$) are vector and axial-vector couplings with
the $Z'$ ($Z$) boson and they are given in Table \ref{tab:table1}. 

\begin{table}
\caption{The family independent vector and axial-vector couplings to new $Z'$
boson predicted by different models. \label{tab:table1} }

\tabcolsep 0.05pt
\small\begin{tabular}{|c|c|c|c|c|c|c|c|}
\hline 
\multicolumn{2}{|c|}{down-type quarks} & \multicolumn{2}{c|}{up-type quarks} & \multicolumn{2}{c|}{charged leptons} & \multicolumn{2}{c|}{neutrinos}\tabularnewline
\hline 
$C{}_{V}^{'}$ & $C{}_{A}^{'}$ & $C{}_{V}^{'}$ & $C{}_{A}^{'}$ & $C{}_{V}^{'}$ & $C{}_{A}^{'}$ & $C{}_{V}^{'}$ & $C{}_{A}^{'}$\tabularnewline
\hline 
\multicolumn{8}{|c|}{$_{Z'_{S}}$}\tabularnewline
\hline 
-$\frac{1}{2}+\frac{2}{3}\sin{}^{2}\theta_{W}$ & -$\frac{1}{2}$ & $\frac{1}{2}-\frac{4}{3}\sin^{2}\theta_{W}$ & $\frac{1}{2}$ & $-\frac{1}{2}+2\sin^{2}\theta_{W}$ & -$\frac{1}{2}$ & $\frac{1}{2}$ & $\frac{1}{2}$\tabularnewline
\hline 
\multicolumn{8}{|c|}{$_{Z'_{\psi}}$}\tabularnewline
\hline 
0 & $\frac{\sqrt{10}}{6}\sin\theta_{W}$ & 0 & $\frac{\sqrt{10}}{6}\sin\theta_{W}$ & 0 & $\frac{\sqrt{10}}{6}\sin\theta_{W}$ & $\frac{\sqrt{10}}{12}\sin\theta_{W}$ & $\frac{\sqrt{10}}{12}\sin\theta_{W}$\tabularnewline
\hline 
\multicolumn{8}{|c|}{$_{Z'_{\chi}}$}\tabularnewline
\hline 
$\frac{\sqrt{6}}{3}\sin\theta_{W}$ & $-\frac{\sqrt{6}}{6}\sin\theta_{W}$ & 0 & $\frac{\sqrt{6}}{6}\sin\theta_{W}$ & $-\frac{\sqrt{6}}{3}\sin\theta_{W}$ & $-\frac{\sqrt{6}}{6}\sin\theta_{W}$ & $-\sqrt{6}\sin\theta_{W}$ & $-\sqrt{6}\sin\theta_{W}$\tabularnewline
\hline 
\multicolumn{8}{|c|}{$_{Z'_{\eta}}$}\tabularnewline
\hline 
$\sin\theta_{W}$ & $\frac{1}{3}\sin\theta_{W}$ & $0$ & $4\sin\theta_{W}$ & $-\sin\theta_{W}$ & $\frac{1}{3}\sin\theta_{W}$ & $-\frac{1}{3}\sin\theta_{W}$ & $-\frac{1}{3}\sin\theta_{W}$\tabularnewline
\hline 
\multicolumn{8}{|c|}{$_{Z'_{B-L}}$}\tabularnewline
\hline 
$\frac{2}{3}$ & $0$ & $\frac{2}{3}$ & $0$ & $-2$ & $0$ & $-1$ & $-1$\tabularnewline
\hline
\end{tabular}
\end{table}

\begin{figure}
\includegraphics[scale=0.65]{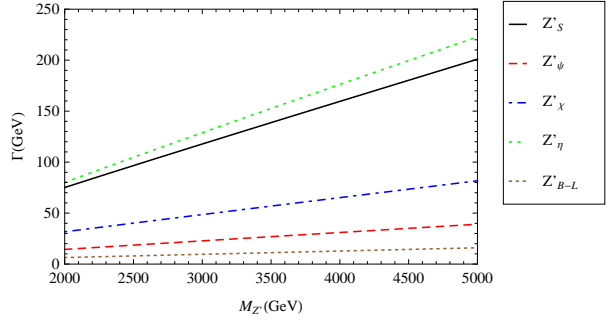}
\caption{The decay widths of $Z'$ boson predicted by different models in the
case of four fermion families. \label{fig:fig1}}
\end{figure}

The decay widths into the heavy fermion pair $F\overline{F}$ and
$W^{+}W^{-}$ bosons are given as

\begin{equation}
\Gamma\left(Z'\rightarrow F\overline{F}\right)=\frac{g_{Z'}^{2}N_{c}}{48\pi M_{Z'}}\sqrt{1-\frac{4M_{F}^{2}}{M_{Z'}^{2}}}\left[(C_{A}^{'F})^{2}\left(-4M_{F}^{2}+M_{Z'}^{2}\right)+(C_{V}^{'F})^{2}\left(M_{Z'}^{2}+2M_{F}^{2}\right)\right]\label{eq:2}
\end{equation}

\begin{eqnarray}
\Gamma\left(Z'\rightarrow W^{+}W^{-}\right) & = & \frac{g_{W}^{2}\cos{}^{2}\theta_{W}\kappa^{2}}{192\pi M_{Z'}M_{W}^{4}}\sqrt{1-\frac{4M_{W}^{2}}{M_{Z'}^{2}}}\left(\frac{M_{Z}}{M_{Z'}}\right)^{4}\nonumber \\
 &  & \times\left[M_{Z'}^{6}+16M_{W}^{2}M_{Z'}^{4}-68M_{W}^{4}M_{Z'}^{2}-48M_{W}^{6}\right]\label{eq:3}
 \end{eqnarray}
where $N_{c}$ is the color factor (3 for quarks and 1 for leptons),
and $g_{Z'}$ is the coupling constant for $Z'$ boson. The $M_{Z}$,
$M_{Z'}$ and $M_{W}$ are the masses for $Z$, $Z'$ and $W$ bosons,
respectively. The $M_{F}$ is the mass of heavy fermion. The mixing
term between the $Z'$ boson and $Z$ boson is assumed to be of the
order of $M_{Z}^{2}/M_{Z'}^{2}$, hence a mixing factor $\kappa$
scales this extension depending on the specific $Z'$ models. In the
sequential model the factor $\kappa$ is chosen to be unity, which
is a reference for the purpose of comparison. In Fig. \ref{fig:fig1}
we present the decay width of $Z'$ boson versus the mass $M_{Z'}$
in case of four fermion families. As it can be seen from
Fig. \ref{fig:fig1} that the total decay width $\Gamma$ for
$Z'_S$, $Z'_\psi$, $Z'_\chi$, $Z'_\eta$ and $Z'_{B-L}$ models are about 
$118$, $23$, $48$, $129$ and $10$ GeV at a mass value of $M_{Z'}=3000$ GeV, respectively. 
For the sequential $Z'$ model, the branching ratios of all fermionic modes 
are not much sensitive to $M_{Z'}$, leading to the fractions of about 0.7, 0.2, 0.1 and
0.02 for $Z'\to q\overline{q}$, $Z'\to\nu\overline{\nu}$, $Z'\to l^{+}l^{-}$
and $Z'\to W^{+}W^{-}$, respectively.
Here, the $q\bar{q}$ mode includes the quarks of four families, and
the $l^{+}l^{-}(\nu\bar{\nu})$ mode includes the charged leptons (neutrinos)
of four families.

\begin{figure}
\includegraphics[scale=0.65]{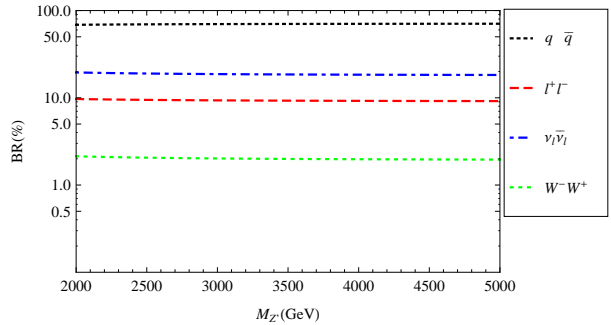}
\caption{The branching ratios of sequential $Z'$ boson into different final states
depending on the mass. \label{fig:fig2}}
\end{figure}

\section{Cross Sections}

Using the interaction Lagrangian (1) we calculate the differential
cross section for the pair production of fourth family quarks and
leptons in the collisions of $e^{+}$ and $e^{-}$ beams. The analytical
expressions for the differential cross section are given in the Appendix.
We calculate the cross section for pair production of fourth family
quarks (leptons) taking their masses in the interval 600-1000 (200-800)
GeV. Table \ref{tab:table2} shows the production cross sections without
$Z'$ contribution at ILC and CLIC energies. The ILC with $\sqrt{s}=1$
TeV has the potential up to the kinematical range ($m_{l',\nu'}\leq500$ GeV)
for the production cross section of fourth family lepton pairs.
However, the CLIC with $\sqrt{s}=3$
TeV extends the mass range for the fourth family fermions. In order
to see the contributions from $Z'$ boson exchange and its interference
we also calculate the cross sections assuming the reference mass values
$M_{b'}=650$ GeV and $M_{\nu'}=100$ GeV with the constraints $M_{b'}-M_{t'}\simeq 50$
GeV and $M_{l'}-M_{\nu'} \simeq 100$ GeV. 
The cross sections for $\ell '$ and $\nu '$ pair production 
through $Z'$ effects are
shown in Table \ref{tab:table3} for ILC with $\sqrt{s}=1$ TeV. 
In Table \ref{tab:table4} the cross sections for pair production of $t'$ and $b'$ quarks 
are presented for CLIC with $\sqrt{s}=3$ TeV through
different $Z'$ models. In Tables \ref{tab:table3} and \ref{tab:table4}, 
we assume the $Z'$ boson mass $m_{Z'}=2500$ GeV.

\begin{table}
\caption{The cross sections (fb) for the fourth family pair production processes
(without $Z'$) at CLIC with $\sqrt{s}=3$ TeV. The numbers in paranthesis
shows the results for ILC with $\sqrt{s}=1$ TeV.\label{tab:table2}}

\begin{tabular}{|c|c|cc|c|c|c|}
\hline 
Mass (GeV) & $e^{+}e^{-}\to\nu'\bar{\nu}'$ & $e^{+}e^{-}\to l'^{+}l'^{-}$ &  & Mass (GeV) & $e^{+}e^{-}\to b'\bar{b}'$ & $e^{+}e^{-}\to t'\bar{t}'$\tabularnewline
\hline
200 & 2.71(22.10) & 12.44(108.60) &  & 600 & 9.17 & 18.80\tabularnewline
300 & 2.67(18.28) & 12.39(100.70) &  & 700 & 8.79 & 18.30\tabularnewline
400 & 2.60(12.66) & 12.31(81.80) &  & 800 & 8.35 & 17.70\tabularnewline
600 & 2.42 & 12.05 &  & 900 & 7.83 & 17.00\tabularnewline
800 & 2.16 & 11.56 &  & 1000 & 7.22 & 16.10\tabularnewline
\hline
\end{tabular}
\end{table}

\begin{table}
\caption{The cross sections for the processes $e^{-}e^{+}\to F\overline{F}$
(where $F=l',\nu'$) at the collision center of mass energy
$\sqrt{s}=1$ TeV. \label{tab:table3}}
\begin{tabular}{|c|c|c|c|c|c|}
\hline 
Cross sections(fb) & $Z'_{S}$ & $Z'_{\psi}$ & $Z'_{\chi}$ & $Z'_{\eta}$ & $Z'_{B-L}$\tabularnewline
\hline
$e^{-}e^{+}\to l'^{-}l'^{+}$ & 105.17 & 107.65 & 99.92 & 105.50 & 104.78\tabularnewline
\hline 
$e^{-}e^{+}\to\nu'\bar{\nu}'$ & 16.04 & 25.55 & 27.03 & 24.18 & 24.48\tabularnewline
\hline
\end{tabular}
\end{table}

\begin{table}
\caption{The cross sections for the process $e^{-}e^{+}\to F\overline{F}$
(where $F=t',b',l',\nu'$) at $\sqrt{s}=3$ TeV. \label{tab:table4}}
\begin{tabular}{|c|c|c|c|c|c|}
\hline 
Cross section (fb) & $Z'_{S}$ & $Z'_{\psi}$ & $Z'_{\chi}$ & $Z'_{\eta}$ & $Z'_{B-L}$\tabularnewline
\hline
\hline 
$e^{-}e^{+}\to t'\bar{t}'$ & 84.75 & 15.72 & 26.46 & 45.74 & 24.14\tabularnewline
\hline 
$e^{-}e^{+}\to b'\bar{b}'$ & 95.13 & 16.16 & 14.91 & 7.48 & 6.84\tabularnewline
\hline 
$e^{-}e^{+}\to l'^{-}l'^{+}$ & 35.59 & 15.62 & 38.51 & 19.35 & 21.05\tabularnewline
\hline 
$e^{-}e^{+}\to\nu'\bar{\nu}'$ & 49.60 & 0.93 & 5.80 & 3.29 & 3.16\tabularnewline
\hline
\end{tabular}
\end{table}

The leptonic decay mode of the $Z'$ boson has lower branching ratio
than the hadronic one, but the cross section for the process $e^{-}e^{+}\to l'^{-}l'^{+}$
is comparable with the $t'\overline{t}'$ pair production for some
$Z'$ models. The intermediate goals after the discovery of the $Z'$
boson and the fourth family fermions would be to understand their
properties and couplings. The forward-backward asymmetry and the invariant
mass spectrum of the heavy fermions could help to identify the nature
of these new particles. 

\section{Forward-Backward Asymmetry}

The forward-backward asymmetry $A_{FB}$ is defined as the relative
difference between the cross sections with $\cos\theta>0$ and $\cos\theta<0$,
being $\theta$ the angle between the heavy fermion $F$ and initial
electron in the center of mass frame:

\[
A_{FB}=\frac{\sigma(\cos\theta>0)-\sigma(\cos\theta<0)}{\sigma(\cos\theta>0)+\sigma(\cos\theta<0)}
\]

\begin{figure}
\includegraphics[scale=0.65]{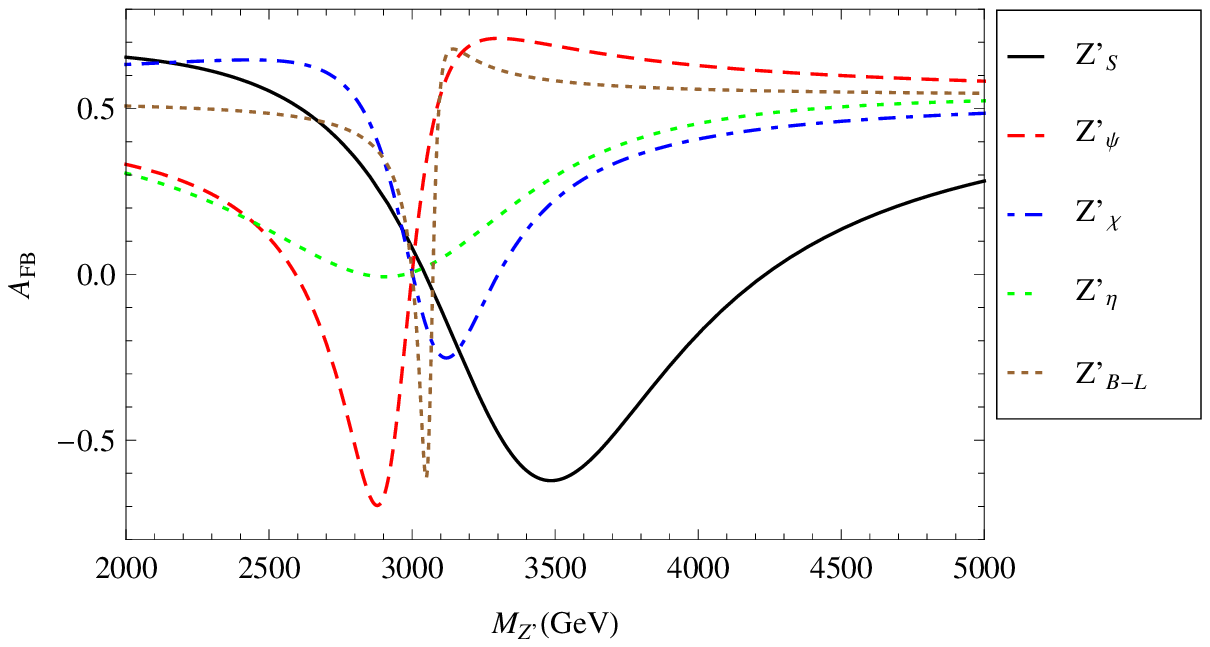}\includegraphics[scale=0.65]{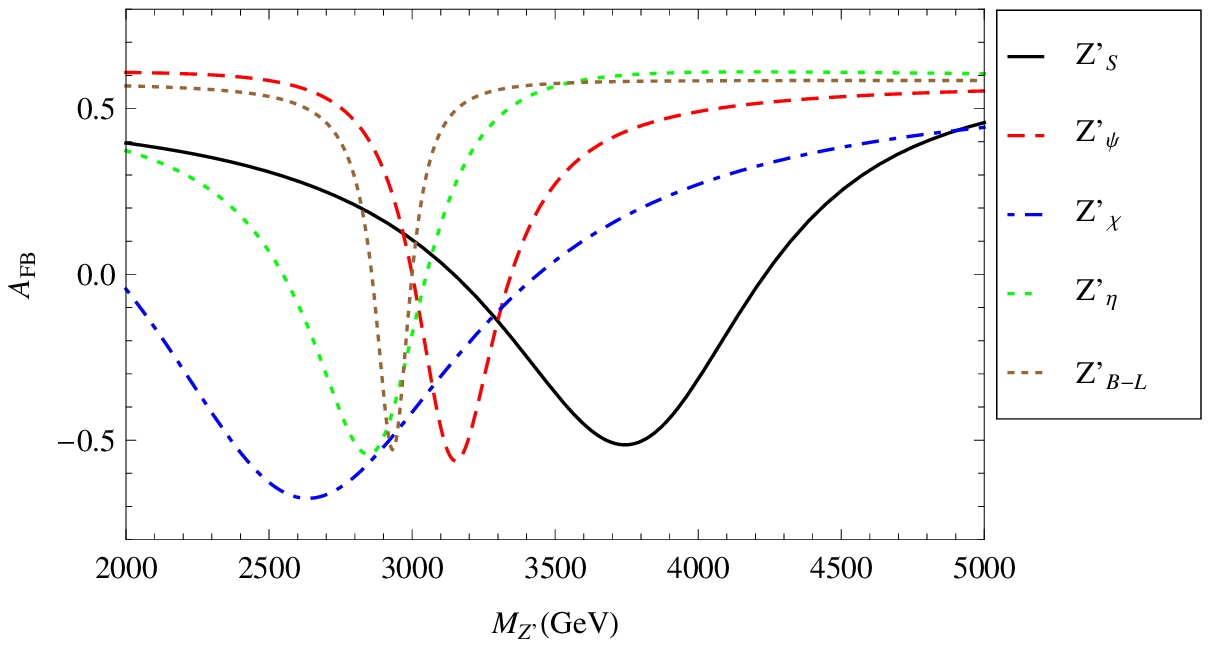}
\caption{Forward-backward asymmetry for $t'$ (left) and $b'$ (right) production within different $Z'$
models at the center of mass energy of 3 TeV.
\label{fig:fig3}}
\end{figure}

\begin{figure}
\includegraphics[scale=0.65]{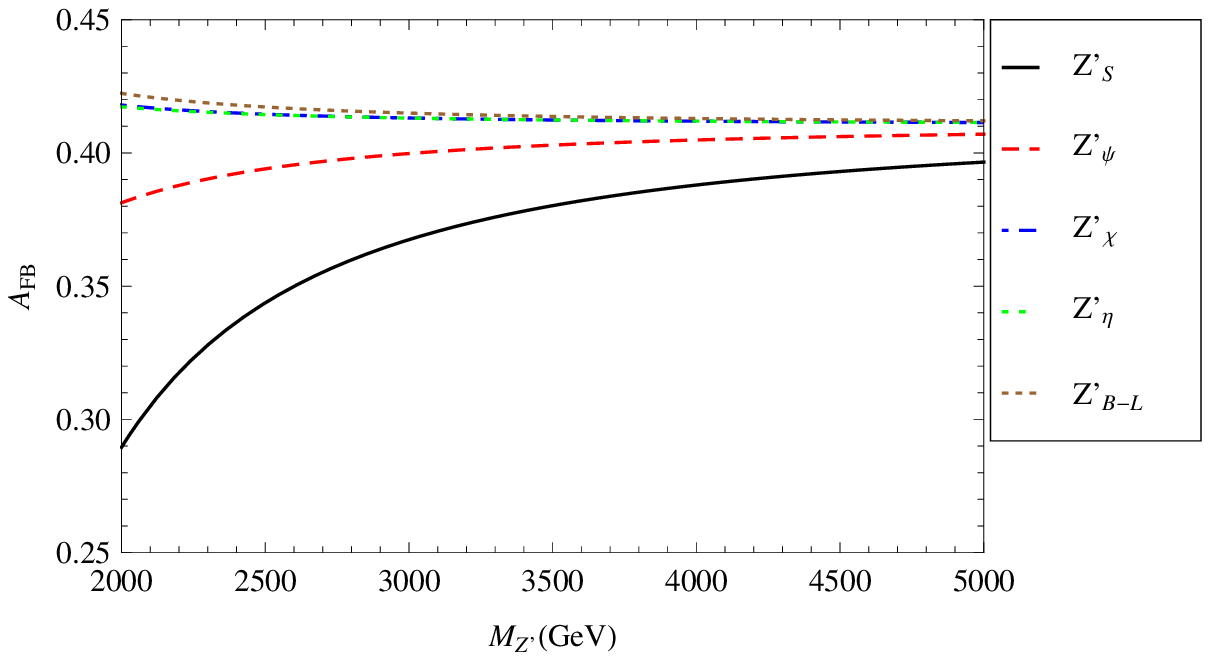}\includegraphics[scale=0.65]{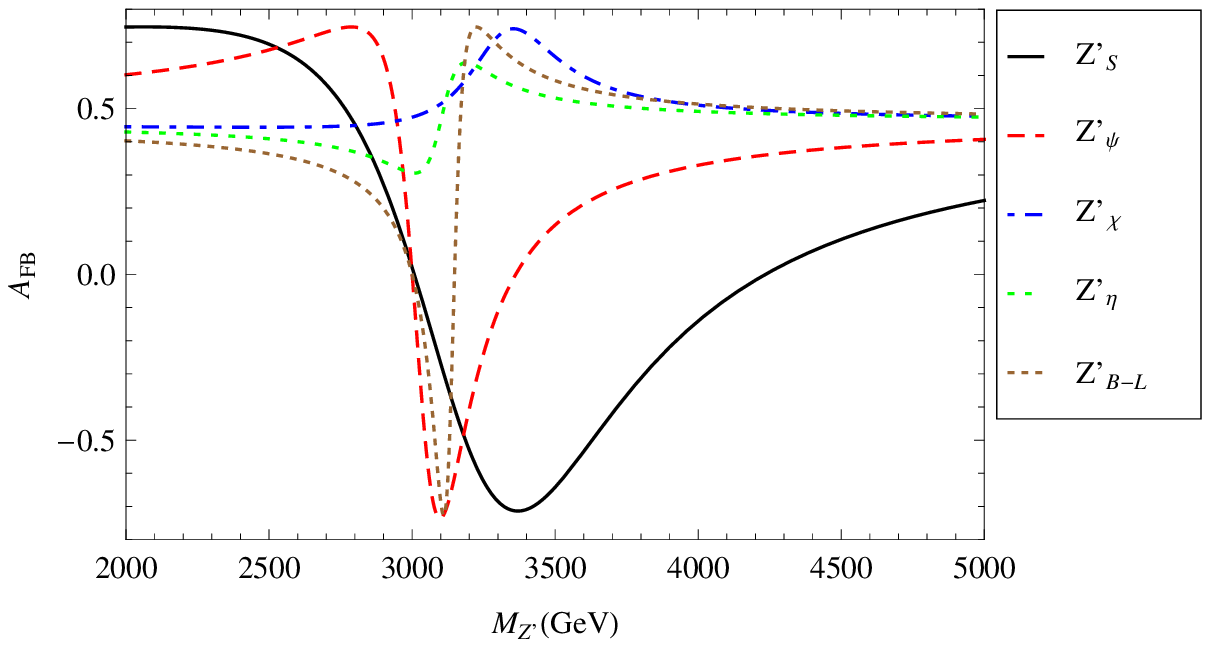}
\caption{Forward-backward asymmetry for $l'$ lepton within different $Z'$
models at the center of mass energies 1 TeV (left) and 3 TeV (right).
\label{fig:fig4}}
\end{figure}

\begin{figure}
\includegraphics[scale=0.65]{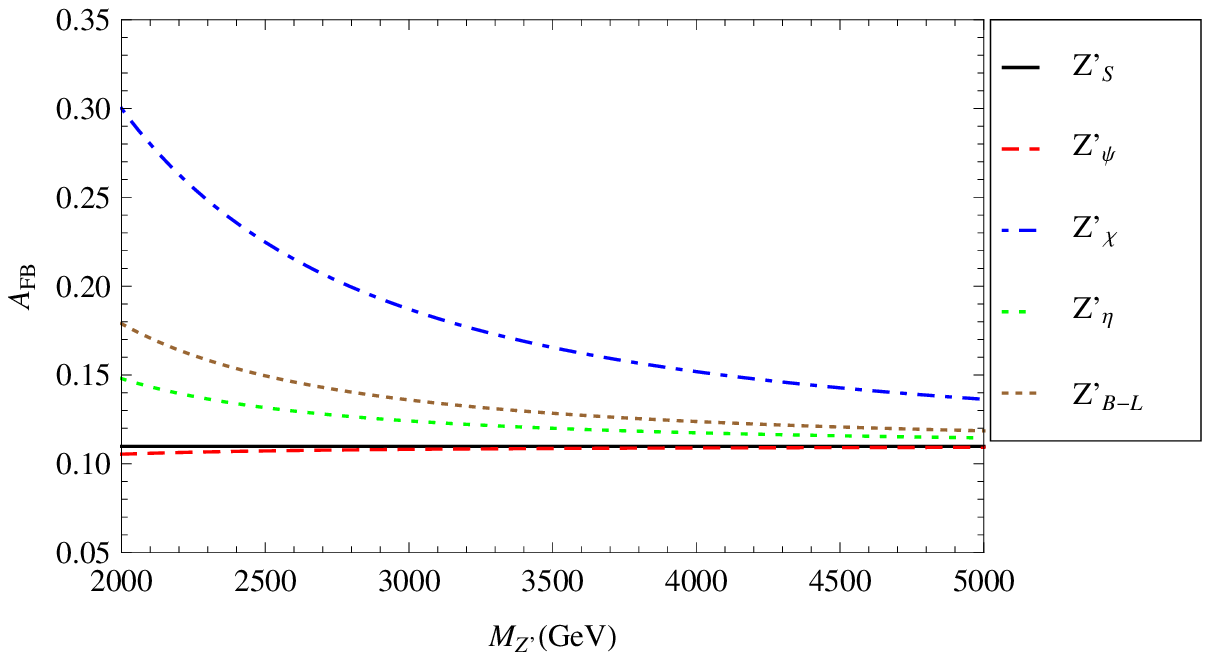}\includegraphics[scale=0.65]{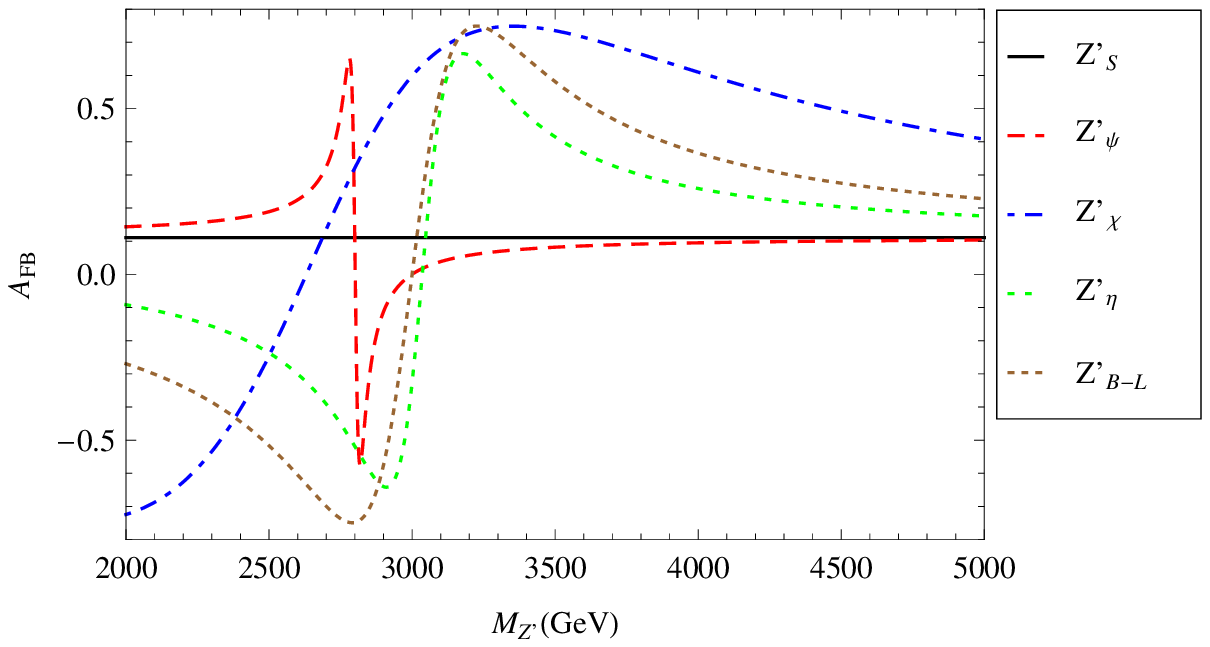}
\caption{The same as Fig. \ref{fig:fig4}, but for $\nu'$. \label{fig:fig5}}
\end{figure}

Since the photon has only vector-like couplings to charged fermions
the photon exchange can not generate an asymmetry, while the $Z$
boson and $Z'$ boson exchange and their interference can generate
asymmetry for the fourth family fermions. If the heavy fermions are
localized differently along a new dynamical symmetry breaking, one
can then expect that the interactions of heavy fermions can be different
from the ones of the light fermions. The presence of the $s$-channel
resonance in $F\overline{F}$ production could be identified by an
examination of the invariant mass distributions with sufficient statistics.
It is seen from Fig. \ref{fig:fig3} that the
asymmetry changes sign at a value of the $M_{Z'}$ near the center
of mass energy. At relatively low $M_{Z'}$ the asymmetry value is around 0.6 (0.4)
for $t'$ ( $b'$) pair production with the contribution of sequential $Z'$
boson at the center of mass energy of 3 TeV. However, it has the
value around 0.3 (0.45) for the large $M_{Z'}$ region.
Figs. \ref{fig:fig4} and \ref{fig:fig5} show the
asymmetries for the fourth family leptons depending on the $Z'$ mass.
Concerning the other $Z'$ models different asymmetry behaviour of the fourth family 
fermions can be seen in Figs. \ref{fig:fig3}-\ref{fig:fig5} depending on the mass of $Z'$ boson.

\begin{figure}
\includegraphics[scale=0.65]{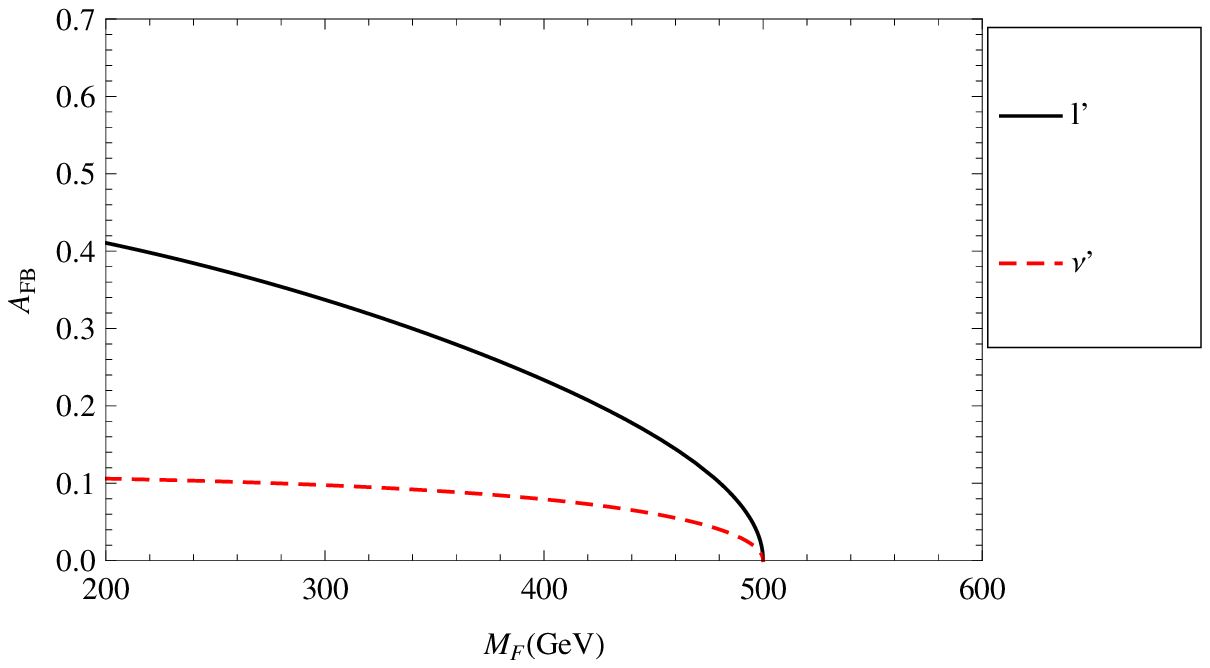}\includegraphics[scale=0.65]{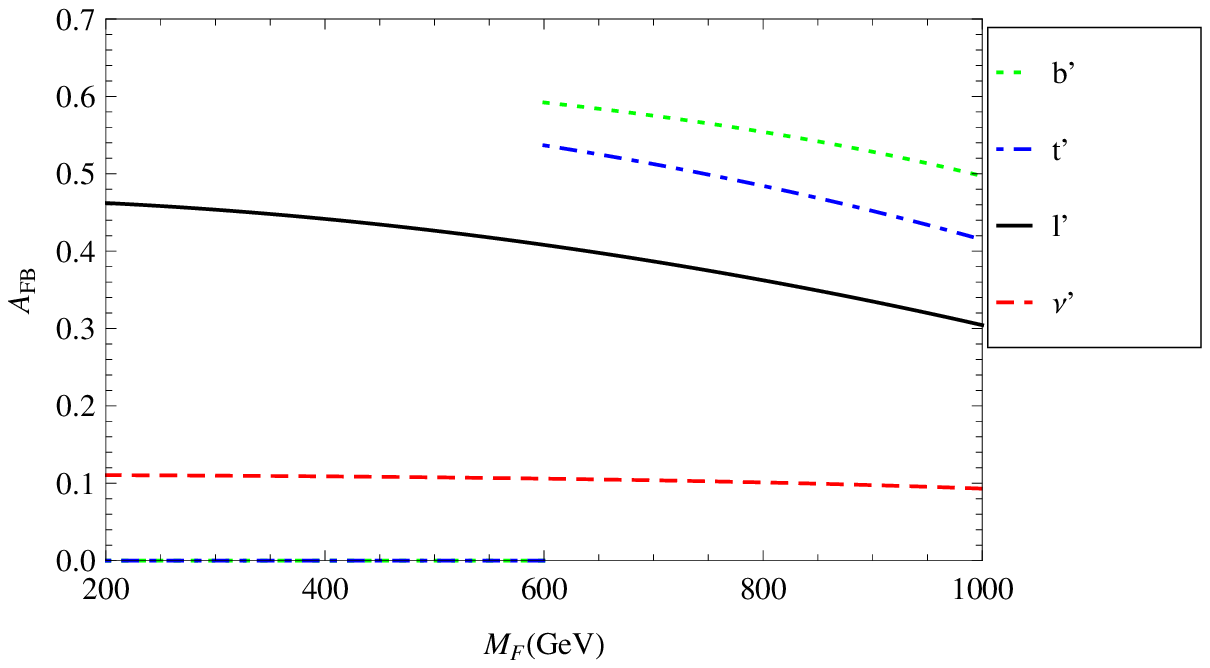}
\caption{The forward-backward asymmetries for fourth family leptons ($l'$, $\nu'$) at $\sqrt{s}=1$ TeV (left) 
and for fourth family fermions ($t'$, $b'$, $l'$, $\nu'$) at $\sqrt{s}=3$ TeV (right) 
depending on their masses. \label{fig:fig6} }
\end{figure}

In order to see how the asymmetry changes depending on the heavy fermion
mass $M_{F}$ we plot Fig. \ref{fig:fig6} without $Z'$ contribution. 
One may comment that the asymmetry $A_{FB}$ for $l'$ pair production is 
higher than that of the $\nu'$ at ILC and CLIC energies. As it can be seen from the differential cross 
section in the Appendix, fourth family fermions have different vector and axial 
couplings to the $Z$ boson. This can generate different asymmetries 
for the pair production of fourth family fermions at linear colliders. 
Pairs of heavy charged fermions can couple to photon and $Z$ 
boson and interfere with each other while the
heavy neutrino pair can couple to $Z$ boson, 
therefore we expect a lower asymmetry for the neutrinos in 
the SM framework.
Considering the current mass limits for $m_{b'}$ and $m_{t'}$ the asymmetry is given 
for the CLIC energy.
For the forward-backward asymmetry of the fourth family fermions depending
on the invariant mass ($M_{F\overline{F}}$) and the initial state
radiation (ISR) and beamstrahlung (BS) effects, we use CalcHEP \cite{Pukhov99}
with the beam parameters for the ILC \cite{Brau07} and CLIC \cite{Assmann00}
as presented in Table \ref{tab:table5}.

\begin{table}
\caption{The collider beam parameters of the ILC and CLIC needed to calculate
the ISR and BS. \label{tab:table5}}

\begin{tabular}{|c|c|c|}
\hline 
 & ILC ($1$ TeV) & CLIC ($3$ TeV)\tabularnewline
\hline
Horizontal beam size (nm) & $640$ & $45$\tabularnewline
\hline 
Vertical beam size (nm) & $5.7$ & $1$\tabularnewline
\hline 
Bunch length (mm) & $0.3$ & $0.044$\tabularnewline
\hline 
Number of particles in the bunch (N) & $2\times10^{10}$ & $3.72\times10^{9}$\tabularnewline
\hline 
Design luminosity (cm$^{-2}$s$^{-1}$) & $2\times10^{34}$ & $5.9\times10^{34}$\tabularnewline
\hline
\end{tabular}
\end{table}

\begin{figure}
\includegraphics[scale=0.6]{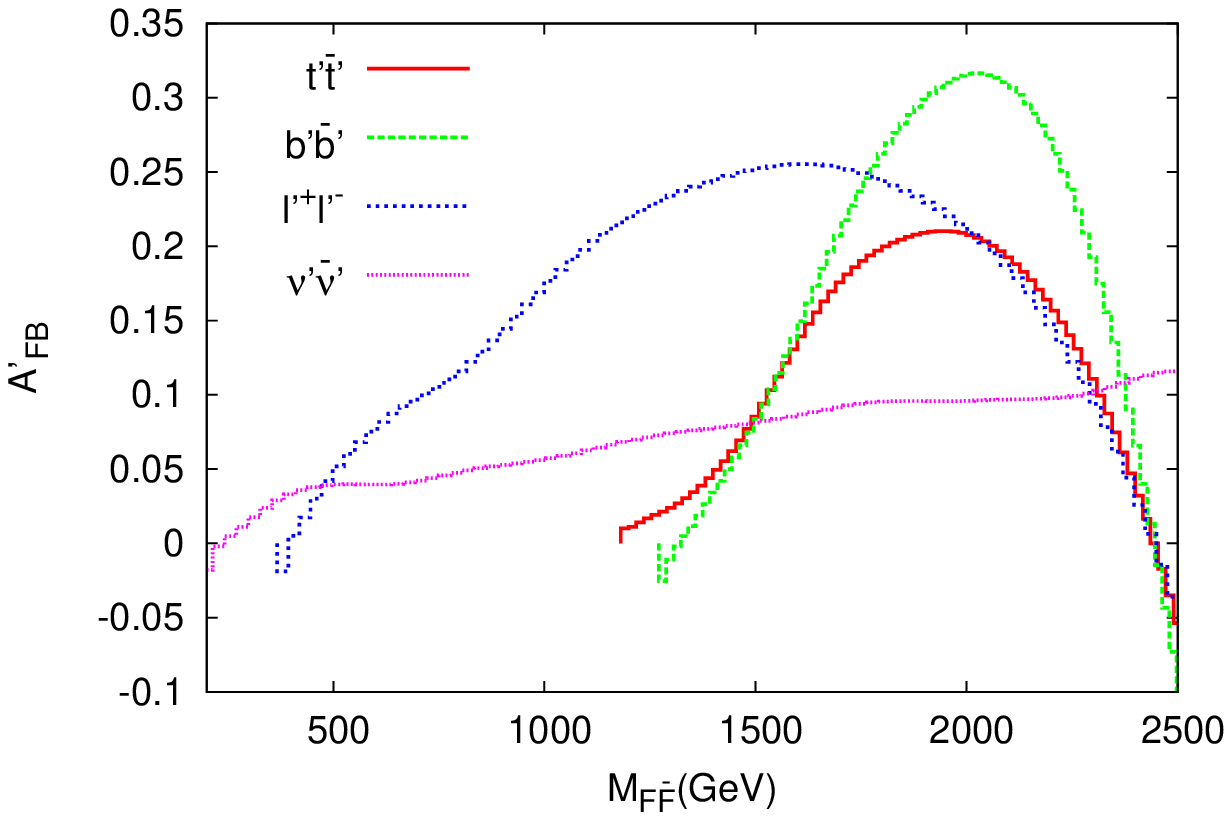} \includegraphics[scale=0.6]{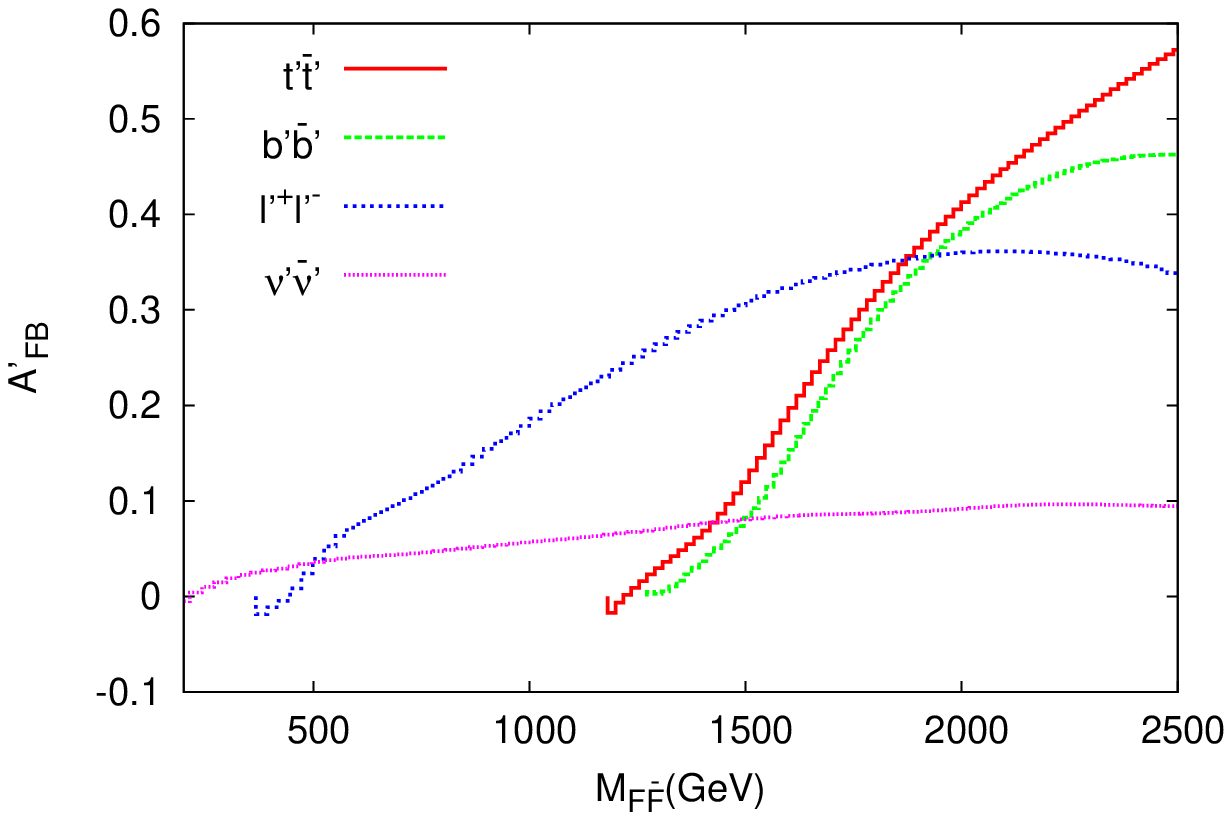}
\caption{Asymmetry depending on the heavy fermion invariant mass for sequential
$Z'_{S}$ model (left) and $Z'_{\psi}$ model (right) for $m_{Z'}=3500$
GeV at CLIC with $\sqrt{s}=3$ TeV. \label{fig:fig7}}
\end{figure}

The asymmetries ($A'_{FB}$) defined in terms of differential cross
sections are presented in Fig. \ref{fig:fig7} depending on the invariant
mass of heavy fermions for $M_{Z'}=3500$ GeV at CLIC with $\sqrt{s}=3$
TeV. For the sequential $Z'$ model, one obtains an asymmetry of 0.3 around 
$M_{b'\bar{b'}}=2000$ GeV at CLIC.
One may compare the distributions between the sequential $Z'$
model and $Z'_{\psi}$ model. It is seen that $Z'_{\psi}$ model generates
more asymmetry for charged fermions depending on the invariant mass
$M_{F\bar{F}}$. One should note that there is a threshold value for
the invariant mass distributions of each type of fermions. These thresholds 
depend on the value of the heavy fermion
mass. The asymmetry for the neutrino remains approximately at the same level
for the interested mass region within these $Z'$ models.

\section{ANALYSIS}

In order to analyze the $Z'$ models we define a $\chi^{2}$ function
given by

\[
\chi^{2}=\frac{(\sigma^{\mbox{with }Z'}-\sigma^{\mbox{no }Z'})^{2}}{\sigma^{\mbox{no }Z'}/(BR\mbox{ }\epsilon\mbox{ }L_{int})}
\]
where $\sigma^{\mbox{with }Z'}$ and $\sigma^{\mbox{no }Z'}$ are
the cross sections for pair production of the fourth family fermions
with a $Z'$ boson and without $Z'$ boson, respectively. The integrated
luminosity $L_{int}$ is taken as 200 fb$^{-1}$ at the center of
mass energy $\sqrt{s}=1$ TeV for ILC and 600 fb$^{-1}$ at $\sqrt{s}=3$
TeV for CLIC. The $BR$ and $\epsilon$ correspond to the branching ratio and
efficiency for considered decay mode, respectively. 

\begin{figure}
\includegraphics[scale=0.6]{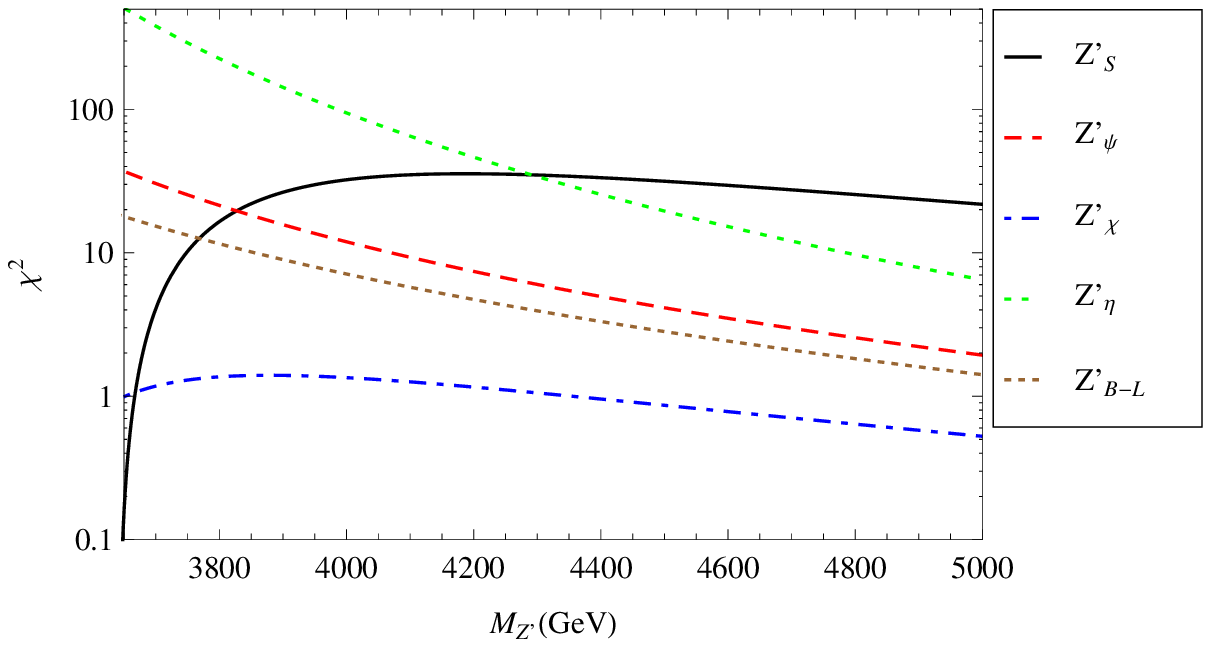} \includegraphics[scale=0.6]{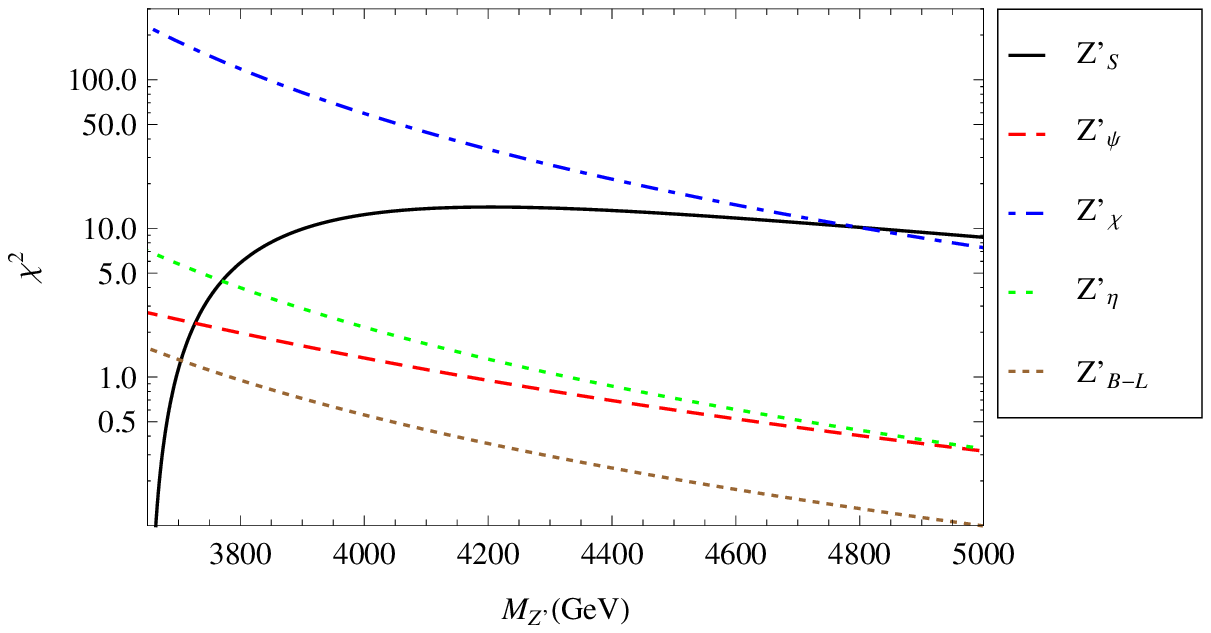}
\caption{The $\chi^{2}$ distribution for $t'$ (left) and $b'$ (right) pair production process depending
on the $Z'$ mass at $\sqrt{s}=3$ TeV. \label{fig:fig8}}
\end{figure}

First, we take into account the pair production of fourth family quarks
($t'$ and $b'$) and their decays via $t'\bar{t}'\to W^{+}bW^{-}\bar{b}$
and $b'\bar{b}'\to W^{-}tW^{+}\bar{t}\to W^{-}W^{+}bW^{+}W^{-}\bar{b}$,
respectively. For $t'$ pair production process, we consider the leptonic
decay of one $W$ boson and hadronic decay of the other $W$ boson
giving the signal $l^{\pm}+2b_{jet}+2j+\mbox{MET}$. In Fig. \ref{fig:fig8},
we plot the $\chi^{2}$ distribution versus $M_{Z'}$ assuming the
mass value $m_{t'}=600$ GeV and CKM4 elements $|V_{t'b'}|=0.993$,
$|V_{t'b}|=0.115$, $|V_{t's}|=0.034$, $|V_{t'd}|=0.006$, $|V_{tb'}|=0.115$,
$|V_{cb'}|=0.034$, $|V_{ub'}|=0.014$ \cite{Eilam09} at the linear collider
center of mass energy $\sqrt{s}=3$ TeV. For $b'$ pair
production, we take into account the same sign $W$ bosons decay leptonically,
while the others decay hadronically, leading to the signal $2l^{\pm}+2b_{jet}+4j+\mbox{MET}$.

For $t'$ ($b'$) pair production we can identify the $Z'_{S}$ model
in the mass range $m_{Z'}=3700-5000$ GeV at linear collider energy of $\sqrt{s}=3$ TeV. 
However, the other models can be identified in the smaller mass
range well above the experimental limits. 
Specifically, for the $b'$ pair production the $Z'_{\chi}$
model will give its signature up to a higher mass value. 

\begin{figure}
\includegraphics[scale=0.6]{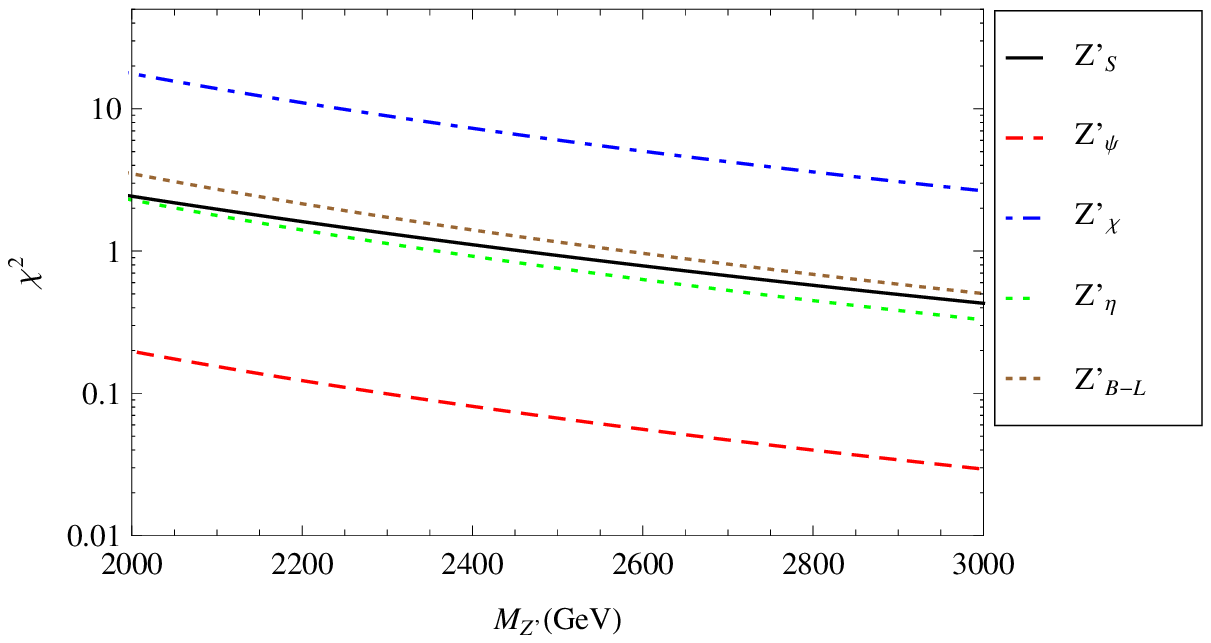} \includegraphics[scale=0.6]{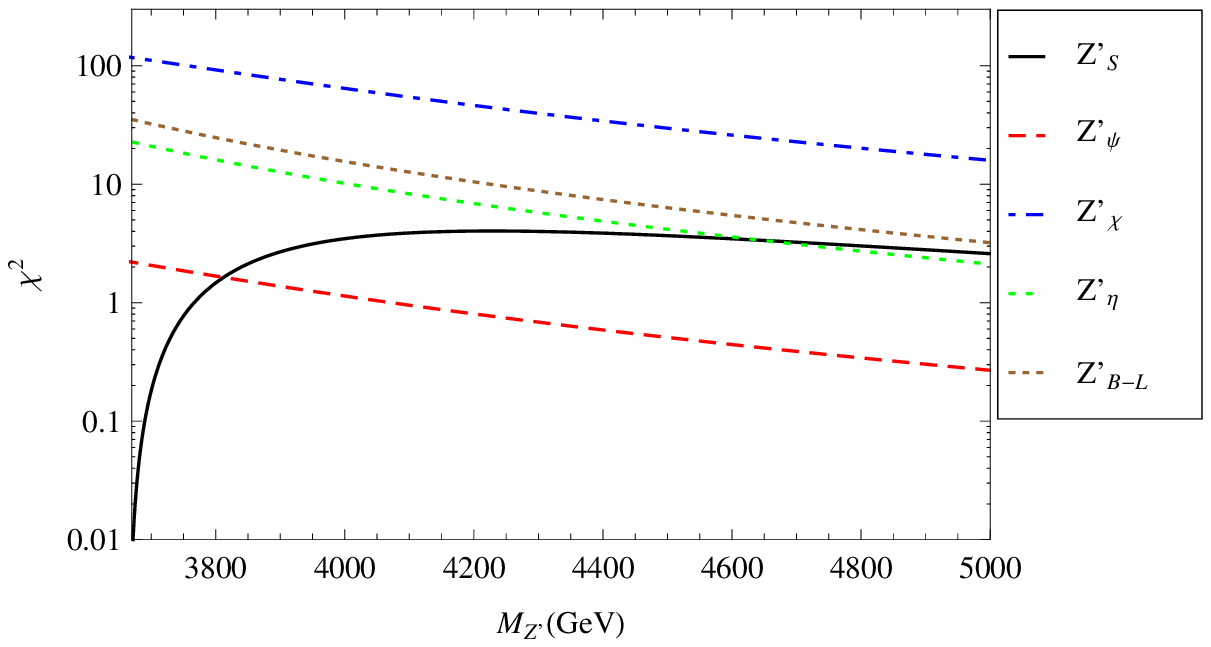}
\caption{The $\chi^{2}$ distribution for $l'$ pair production process depending
on the $Z'$ mass at $\sqrt{s}=1$ TeV (left) and $\sqrt{s}=3$ TeV
(right). \label{fig:fig9}}
\end{figure}

\begin{figure}
\includegraphics[scale=0.6]{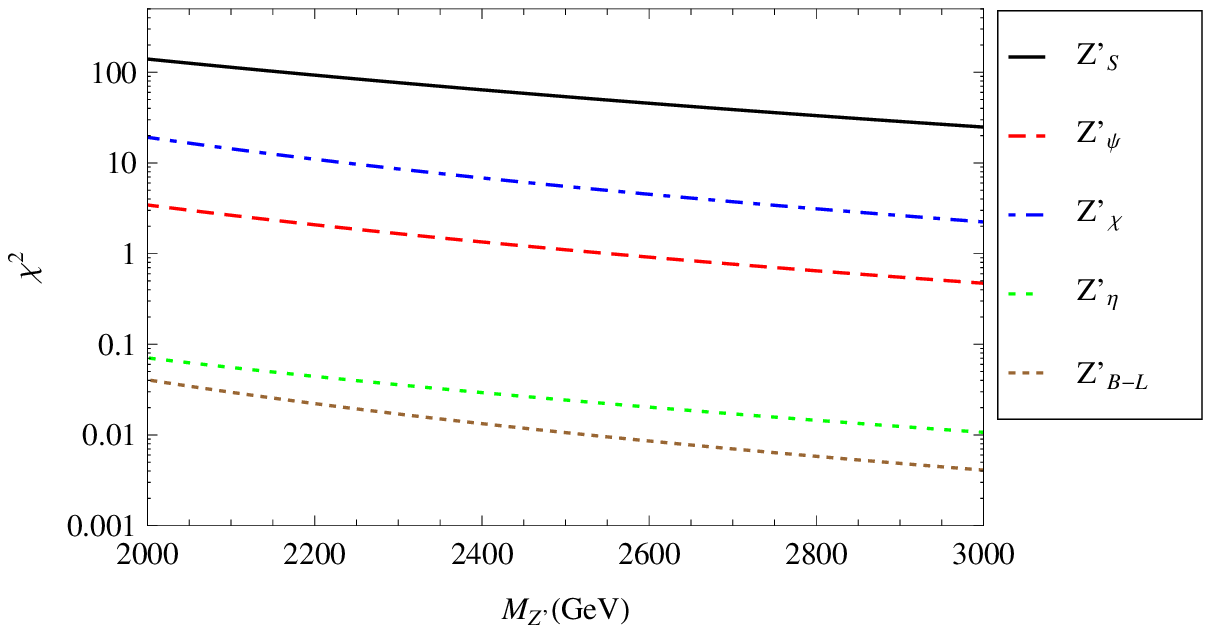} \includegraphics[scale=0.6]{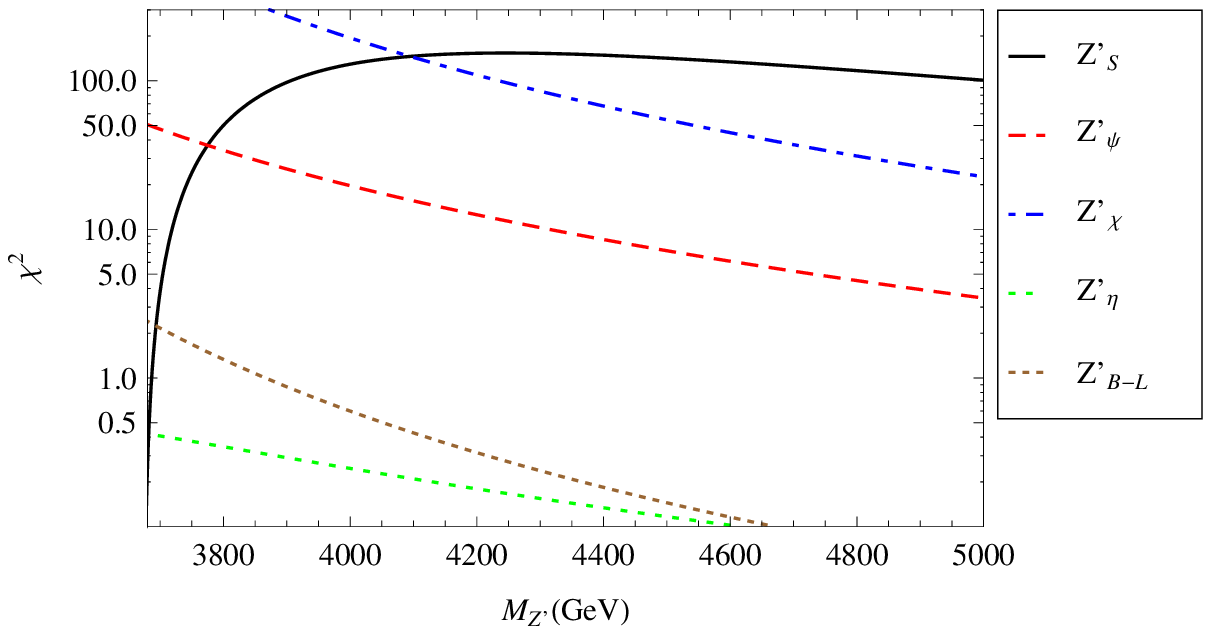}
\caption{The same as Fig. \ref{fig:fig9}, but for the $\nu'$ pair production
at $\sqrt{s}=1$ TeV (left) and $\sqrt{s}=3$ TeV (right). \label{fig:fig10}}
\end{figure}

Second, we consider the pair production of fourth family charged lepton
and neutrino ($l'$ and $\nu'$) and their decays 
via $l'\bar{l}'\to W^{-}\nu'W^{+}\bar{\nu'}\to W^{-}\mu^{\mp}W^{\pm}W^{+}\mu^{\mp}W^{\pm}$
and $\nu'\bar{\nu}'\to W^{\pm}\mu^{\mp}W^{\pm}\mu^{\mp}$ assuming
the Majorana nature of the neutrino, respectively. For $l'$ pair
production, we assume three same sign $W$ bosons decay leptonically,
while the other decays hadronically, giving the signal $3l^{\pm}+2\mu^{\mp}+2j+\mbox{MET}$.
The results for this signal are given in Table ~\ref{tab:table6} and ~\ref{tab:table7}. 
However, we also consider the final state $l'\bar{l}'\to W^{-}\nu'W^{+}\bar{\nu'}\to 8j+2\mu^\pm$ 
which is more convenient to separate $Z'$ 
models at linear colliders for one year of operation. 
In Fig. \ref{fig:fig9}, we plot the $\chi^{2}$ distribution for the $2\mu^\pm+8j$ signal 
versus $M_{Z'}$ assuming the mass value $m_{l'}=200$ GeV and PMNS4 elements
$U_{\nu'l'}>0.996$ and $U_{\nu'l}<0.092$ \cite{Schmidt11} at the
collider energies $\sqrt{s}=1$ TeV and 3 TeV. For $\nu'$ pair production, 
we assume the $W$ bosons decay hadronically, leading to the
signal $2\mu^{\pm}+4j$. The $\chi^{2}$
distribution is given in Fig. \ref{fig:fig10} depending on the $M_{Z'}$ 
assuming the value $m_{\nu'}=100$
GeV at the center of mass energies $\sqrt{s}=1$ TeV and 3 TeV.

In the $l'$ pair and $\nu'$ pair search we can identify the $Z'_{\chi}$
model in the mass range $2000<m_{Z'}<2700$ GeV at $\sqrt{s}=1$ TeV. A higher
center of mass energy $\sqrt{s}=3$ TeV expands the accessibility 
to the range $m_{Z'}=3700-5000$ GeV.

The number of signal events for the fourth family pair production
processes at the center of mass energies $\sqrt{s}=1$ TeV and 3 TeV
are given in Table \ref{tab:table6} and \ref{tab:table7}, respectively.
Here, we take the mass of fourth family $b'$ quark as $m_{b'}=650$ GeV and 
fourth family lepton mass as $m_{l'}=200$ GeV. The corresponding background
events are also given in the last column of Table \ref{tab:table6}
and \ref{tab:table7} for 200 fb$^{-1}$ and 600 fb$^{-1}$, respectively.
When calculating the number of signal and background events we take
into account the corresponding branching ratios and the efficiency
factors for the given channel of signal. In the final state including
$b$-quarks we take the $b$-tagging efficiency as $\epsilon=0.5$.

\begin{table}
\caption{Number of signal ($l'$ and $\nu '$ pairs) and background events for relevant final states at
$\sqrt{s}=1$ TeV and $L_{int}=200$ fb$^{-1}$. The numbers in the
paranthesis denote corresponding signal significances. \label{tab:table6}}
\tiny\begin{tabular}{|l|c|c|c|c|c|l|}
\hline
Signal & $Z'_{S}$ & $Z'_{\psi}$ & $Z'_{\chi}$ & $Z'_{\eta}$ & $Z'_{B-L}$ & Background\tabularnewline
\hline
$l'\overline{l}'\rightarrow2\mu^{-}+3\mu^{+}+2j+\mbox{MET}$ & 15.6(110.31) & 15.9(112.85) & 14.8(104.79) & 15.6(110.59) & 15.5(109.88) & $\begin{array}{l}
ZZW^{+}W^{-}\\
0.02\end{array}$
\tabularnewline
\hline
$\nu'\overline{\nu}'\rightarrow2\mu^{-}+4j$ & 151.9(159.29) & 241.9(253.66) & 255.9(268.29) & 229.0(240.09) & 231.8(243.00) & $\begin{array}{l}
W^{+}W^{-}W^{+}W^{-}\\
0.91\end{array}$\tabularnewline
\hline
\end{tabular}
\end{table}

\begin{table}
\caption{The number of signal ($t'$, $b'$, $l'$ and $\nu '$ pairs) and background events for $\sqrt{s}=3$ TeV and
$L_{int}=600$ fb$^{-1}$. The numbers in the paranthesis denote the significances. \label{tab:table7}}
\tiny\begin{tabular}{|l|c|c|c|c|c|l|}
\hline 
Signal & $Z'_{S}$ & $Z'_{\psi}$ & $Z'_{\chi}$ & $Z'_{\eta}$ & $Z'_{B-L}$ & Background\tabularnewline
\hline 
$t'\overline{t}'\rightarrow l^{+}+2b_{j}+2j+\mbox{MET}$ & 1582.8(62.15) & 293.6(11.53) & 494.2(19.41) & 854.3(33.55) & 450.8(17.70) & $\begin{array}{l}
W^{+}bW^{-}\bar{b}\\
648.5\end{array}$\tabularnewline
\hline 
$b'\overline{b}'\rightarrow2l^{+}+2b_{j}+4j+\mbox{MET}$ & 206.0(193.82) & 34.9(32.92) & 32.3(30.39) & 16.2(15.23) & 14.8(13.93) & $\begin{array}{l}
W^{-}tW^{+}\bar{t}\\
1.13\end{array}$\tabularnewline
\hline 
$l'\overline{l}'\rightarrow2\mu^{-}+3\mu^{+}+2j+\mbox{MET}$ & 15.6(44.95) & 6.8(19.72) & 16.9(48.64) & 8.5(24.45) & 9.2(26.59) & $\begin{array}{l}
ZZW^{+}W^{-}\\
0.12\end{array}$\tabularnewline
\hline 
$\nu'\overline{\nu}'\rightarrow2\mu^{-}+4j$ & 1385.5(624.02) & 26.0(11.72) & 162.0(72.96) & 91.7(41.32) & 88.3(39.76) & $\begin{array}{l}
W^{+}W^{-}W^{+}W^{-}\\
4.93\end{array}$\tabularnewline
\hline
\end{tabular}
\end{table}

In order to estimate signal significance for the production of fourth
family fermions we use signal and background events at linear colliders
with $\sqrt{s}=1$ TeV and 3 TeV. For an illustration, considering 
$2\mu^- +3\mu^+ +2j+MET$ signal 
we have the signal significances
($S/\sqrt{B})$ in the framework of $Z'_{S}$ model as 110.3 (44.95) and 
159.3 (624.02) for the $l'$
and $\nu'$ pair production at $1$ ($3$) TeV, respectively. 
We will have more number of signal events if we choose the channel 
$2\mu^\pm+8j$ for the $l'$ pair production. For this channel we expect 
quite low background due to at least six gauge bosons in the final state.
At the ILC (CLIC), we obtain 460.5 (459.7), 471.4 (201.7), 437.5 (497.5), 461.9 (250.0) and 458.8 (271.8) 
events for the signal $2\mu^\pm+8j$ within 
the $Z'_S$, $Z'_\psi$, $Z'_\chi$, $Z'_\eta$ and $Z'_{B-L}$ models, respectively.
Concerning the fourth family quarks 
$t'$ and $b'$ we have the significances $62.15$ and $193.82$ for $Z'_S$ model at $\sqrt{s}=3$ TeV.
Providing the fourth family
fermions exist within the considered mass range, the $Z'$ models
can be probed with a large significance at linear colliders.

\section{Conclusions}

We emphasize that exploring the $F\overline{F}$ production cross
sections and forward-backward asymmetry at linear colliders will allow
further tests of the new models beyond the SM. Taking the masses of
fourth family fermions as $m_{t'}=600$ GeV and $m_{l'}=200$ GeV
with the constraints $m_{b'}-m_{t'}=50$ GeV and $m_{l'}-m_{\nu'}=100$
GeV, the CLIC (ILC) can produce fourth family fermions $t'$, $b'$,
$l'$ and $\nu'$ signal events $1582$, $206$, $15$ ($15$), $1385$ ($151$) per
year, respectively. We study the dependence of $A_{FB}^{F}$ on the
heavy fermion invariant mass $M_{F\overline{F}}$. At CLIC (ILC),
the forward backward asymmetry for $t'$, $b'$, $l'$ and $\nu'$ pair production 
without $Z'$ contribution can be calculated as 0.55, 0.60,
0.45 (0.40) and 0.10 (0.10) at their mass bounds, respectively. These asymmetries 
can be affected with the $Z'$ masses in
the framework of models. For an invariant mass of $M_{F\bar{F}}= 2$
TeV, the $t'$ and $b'$ quarks can produce a forward backward asymmetry
of $A'_{FB}\simeq 0.4$ if the $Z'_{\psi}$ model is realized. However,
heavy charged lepton FB asymmetry can also be measured at relatively
low invariant mass range. Performing a $\chi^{2}$ analysis using
the cross sections we have the mass range for the $Z'$ boson which
can be accessible at the linear collider experiments. We found that
the $Z'$ models give different predictions for the observables and
their correlations, and they may be distinguished by jointly studying
these observables at linear colliders.

\section*{APPENDIX}

The differential cross section for the process $e^{+}e^{-}\to F\overline{F}$
is given by

\begin{eqnarray*}
\frac{d\sigma\left(e^{-}e^{+}\rightarrow F\overline{F}\right)}{dt} & = & \frac{1}{16\pi s^{2}}\left\{ \frac{2g_{e}^{4}}{s^{2}}\left[A_{1}Q_{F}^{2}\right]+\frac{g_{Z}^{4}}{8\left[\left(M_{Z}^{2}-s\right)^{2}+M_{Z}^{2}\Gamma_{Z}^{2}\right]}\right.\\
 & \times & \left[\left(C_{A}^{e^{2}}+C_{V}^{e^{2}}\right)\left(C_{V}^{F^{2}}A_{1}+C_{A}^{F^{2}}A_{2}\right)+4C_{A}^{e}C_{A}^{F}C_{V}^{e}C_{V}^{F}sA_{3}\right]\\
 & + & \frac{g_{Z'}^{4}}{8\left[\left(M_{Z'}^{2}-s\right)^{2}+M_{Z'}^{2}\Gamma_{Z'}^{2}\right]}\left[\left(C_{A}^{'e{}^{2}}+C_{V}^{'e{}^{2}}\right)\left(C_{V}^{'F{}^{^{2}}}A_{1}+C_{A}^{'F{}^{^{2}}}A_{2}\right)\right.\\
 & + & \left.4C_{A}^{'e}C_{A}^{'F}C_{V}^{'e}C_{V}^{'F}sA_{3}\right]\\
 & - & \frac{g_{e}^{2}g_{Z}^{2}\left(-Q_{F}\right)\left(M_{Z}^{2}-s\right)}{2s\left[\left(M_{Z}^{2}-s\right)^{2}+M_{Z}^{2}\Gamma_{Z}^{2}\right]}\left(C_{A}^{e}C_{A}^{F}sA_{3}+C_{V}^{e}C_{V}^{F}A_{1}\right)\\
 & - & \frac{g_{e}^{2}g_{Z'}^{2}\left(-Q_{F}\right)\left(M_{Z'}^{2}-s\right)}{2s\left[\left(M_{Z'}^{2}-s\right)^{2}+M_{Z'}^{2}\Gamma_{Z'}^{2}\right]}\left(C_{A}^{'e}C_{A}^{'F}sA_{3}+C_{V}^{'e}C_{V}^{'F}A_{1}\right)\\
 & + & \frac{g_{Z}^{2}g_{Z'}^{2}\left[M_{Z}^{2}\left(M_{Z'}^{2}-s\right)+M_{Z}M_{Z'}\Gamma_{Z}\Gamma_{Z'}+s\left(s-M_{Z'}^{2}\right)\right]}{\left[\left(M_{Z}^{2}-s\right)^{2}+M_{Z}^{2}\Gamma_{Z}^{2}\right]\left[\left(M_{Z'}^{2}-s\right)^{2}+M_{Z'}^{2}\Gamma_{Z'}^{2}\right]}\\
 & \times & \left\{ C_{A}^{e}\left[C_{A}^{'e}\left(C_{A}^{F}C_{A}^{'F}A_{2}+C_{V}^{F}C_{V}^{'F}A_{1}\right)+C_{V}^{'e}sA_{3}\left(C_{A}^{F}C_{V}^{'F}+C_{A}^{'F}C_{V}^{F}\right)\right]\right.\\
 & + & \left.\left.C_{V}^{e}\left[C_{A}^{F}\left(C_{A}^{'e}C_{V}^{'F}sA_{3}+C_{A}^{'F}C_{V}^{'e}A_{2}\right)+C_{V}^{F}\left(C_{A}^{'e}C_{A}^{'F}sA_{3}+C_{V}^{'e}C_{V}^{'F}A_{1}\right)\right]\right\} \right\} \end{eqnarray*}
where $A_{1}=\left(s+t\right)^{2}+t^{2}-4M_{F}^{2}t+2M_{F}^{4}$,
$A_{2}=A_{1}-4M_{F}^{2}s$ and $A_{3}=s+2t-2M_{F}^{2}$. The $g_{Z}$
and $g_{Z'}$ are the coupling constants of the neutral current interactions
with the gauge bosons $Z$ and $Z'$, respectively. The $C_{V}^{'F}$
($C_{V}^{F}$) and $C_{A}^{'F}$ ($C_{A}^{F}$) are vector and axial-vector
couplings with the $Z'$ ($Z$) boson. The $s$ and $t$ are Mandelstam
variables. $M_{Z}$ and $M_{Z'}$ are the masses of $Z$ and $Z'$
bosons; $M_{F}$ is the heavy fermion mass. $\Gamma_{Z}$ and $\Gamma_{Z'}$
are decay widths for $Z$ and $Z'$ bosons, respectively. In order
to obtain the differential cross section depending on the scattering
angle, $d\sigma/d\cos\theta$, the expression $d\sigma/dt$ should
be multiplied by the factor $s\beta/2$ where $\beta=\sqrt{1-4M_{F}^{2}/s}$,
and here we use $t=M_{F}^{2}-s(1-\beta\cos\theta)$/2.

\begin{acknowledgments}
The work is supported in part by Turkish Atomic Energy Authority (TAEK)
under Grant No. CERN-A5.H2.P1.01-10. The work of O.C. and V.C. is
supported in part by the State Planning Organization (DPT) under Grant
No. DPT2006K-120470.
\end{acknowledgments}

\end{document}